\documentclass[preprint,showpacs,showkeys,preprintnumbers,amsmath,amssymb]{revtex4-1}

\usepackage{graphicx}
\usepackage{dcolumn}
\usepackage{bm}

\begin{document}

\preprint{APS/123-QED}

\title{Specific heat and $\mu$SR study on the noncentrosymmetric superconductor LaRhSi$_{3}$}
\author{V. K. Anand}
\email{vivekkranand@gmail.com}
\author{A. D. Hillier}
\author{D. T. Adroja}
\email{devashibhai.adroja@stfc.ac.uk}
\affiliation{ISIS Facility, Rutherford Appleton Laboratory, Chilton, Didcot Oxon, OX11 0QX, UK}
\author{A. M. Strydom}
\affiliation{Physics Department, University of Johannesburg, PO Box 524, Auckland Park 2006, South Africa}
\author{H. Michor}
\affiliation{Institute of Solid State Physics, Vienna University of Technology, A-1040 Wien, Austria}
\author{K. A. McEwen}
\affiliation{Department of Physics and Astronomy and London Centre for Nanotechnology, University College London, Gower Street, London WC1E 6BT, United Kingdom}
\author{B. D. Rainford}
\affiliation{Physics Department, Southampton University, Southampton SO17 1BJ, UK}
\date{\today}
\begin{abstract}

           We have investigated the superconducting properties of the noncentrosymmetric superconductor LaRhSi$_{3}$ by performing magnetization, specific heat, electrical resistivity and muon spin relaxation ($\mu$SR) measurements. LaRhSi$_{3}$ crystallizes with the BaNiSn$_{3}$-type tetragonal structure (space group \textit{I4 mm}) as confirmed through our neutron diffraction study. Magnetic susceptibility, electrical resistivity and specific heat data reveal a sharp and well defined superconducting transition at $T_{c}$ = 2.16 $\pm$ 0.08 K. The low temperature specific heat data reveal that LaRhSi$_{3}$ is a weakly coupled bulk BCS superconductor  and has an s-wave singlet ground state with an isotropic energy gap of $\sim$ 0.3 meV, $2 \Delta_{0} /k_{B}T_{c}$ = 3.24. The specific heat data measured in applied magnetic field strongly indicate a type-I behaviour. Type-I superconductivity in this compound is also inferred from the Ginzburg-Landau parameter, $\kappa$ = 0.25. Various superconducting parameters, including the electron-phonon coupling strength, penetration depth and coherence length, characterize LaRhSi$_{3}$ as a moderate dirty-limit superconductor. A detailed study of the magnetic field-temperature ($H-T$) phase diagram is presented and from a consideration of the free energy, the thermodynamic critical field, $H_{c0}$ is estimated to be 17.1 $\pm$ 0.1 mT, which is in very good agreement with that estimated from the transverse field $\mu$SR measurement that gives $H_{c0}$ = 17.2 $\pm$ 0.1 mT. The transverse field $\mu$SR results are consistent with conventional type-I superconductivity in this compound. Further, the zero-field $\mu$SR results indicate that time reversal symmetry is preserved when entering the superconducting state, also supporting a singlet pairing superconducting ground state in LaRhSi$_{3}$.

\end{abstract}

\maketitle


\section*{Introduction}

The inversion symmetry of a crystal structure plays a central role in the formation of Cooper pairs in conventional superconductors. Therefore with the advent of superconductivity in CePt$_{3}$Si  \cite{1,2}, which lacks inversion symmetry along the c-axis, noncentrosymmetric superconductors have evolved as a hot topic of current research both from experimental and theoretical points of view. The superconducting ground state of CePt$_{3}$Si presents many unusual features due to the presence of an antisymmetric spin-orbit coupling (ASOC) as a consequence of the lack of inversion symmetry, as is well summarised in Ref. 2. The solid solutions Li$_{2}$(PdPt)$_{3}$B  \cite{3,4} and the intermetallic compounds CeRhSi$_{3}$  \cite{5,6}, CeIrSi$_{3}$ \cite{7}, CeCoGe$_{3}$ \cite{8}, LaNiC$_{2}$  \cite{9,10}, BaPtSi$_{3}$ \cite{11}, T$_{2}$Ga$_{9}$(T=Rh,Ir)  \cite{12,13}, and Mg$_{10}$Ir$_{19}$B$_{16}$  \cite{14} are other major examples of known noncentrosymmetric superconductors. Among these CeRhSi$_{3}$, CeIrSi$_{3}$ and CeCoGe$_{3}$ show superconductivity only under the application of pressure, while others have a superconducting ground state at ambient pressure.

The lack of inversion symmetry leads to a non uniform lattice potential which is sensed by the conduction electrons, resulting in a splitting of spin-up and spin-down energy bands and hence a split Fermi surface. In a conventional superconductor Cooper pairs are formed by two electrons having a symmetric orbital state and an antisymmetric spin state both of which belong to the {\it same} Fermi surface.  In contrast, in noncentrosymmetric superconductors the two electrons forming Cooper pairs belong to two {\it different} Fermi surfaces corresponding to the spin-up and spin-down bands. This makes the physics of superconductivity in noncentrosymmetric systems substantially different from that in centrosymmetric systems to which most of the known superconductors belong. From theoretical considerations, the lack of inversion symmetry introduces an antisymmetric spin-orbit coupling (ASOC) which removes the spin degeneracy of the conduction band electrons and therefore in noncentrosymmetric superconductors the spin and orbital parts of the Cooper pairs cannot be treated independently  \cite{15,16,17,18,19}. Further, parity is no longer a good quantum number and a parity mixing is expected, whereby the Cooper pairs of noncentrosymmetric superconductors may contain an admixture of spin-singlet and spin-triplet states.

The symmetry of the superconducting order parameter is very important for understanding the nature of the superconducting ground state. Both time reversal symmetry and inversion symmetry are critical in determining the parity states. While time reversal invariance provides the necessary conditions for spin-singlet pairing, for spin-triplet pairing inversion symmetry is required additionally. In the absence of inversion symmetry, spin-triplet pairing is forbidden: this leads to a contradictory situation in the noncentrosymmetric heavy fermion superconductor CePt$_{3}$Si where the absence of paramagnetic limiting favours spin-triplet pairing  \cite{2}. Therefore a two-component order parameter consisting of mixed spin-singlet and spin-triplet states seems to be appropriate for CePt$_{3}$Si. However, despite extensive efforts by many condensed matter physicists working on noncentrosymmetric superconductors, very little is known so far about the superconducting order parameter in these systems. For example, key issues such as whether they possess a common unusual pairing symmetry, and, if this is the case, what is the nature of the superconducting-gap symmetry remain unsettled. Further investigations on noncentrosymmetric superconductors are therefore required to address such issues.

The Ce-based noncentrosymmetric superconductors CePt$_{3}$Si, CeRhSi$_{3}$, CeIrSi$_{3}$ and CeCoGe$_{3}$ all are situated close to a magnetic quantum critical point, where the presence of magnetic order and heavy fermion behaviour makes it more complicated to extract the physics of inversion symmetry breaking and superconductivity. Therefore a system that is situated far away from a magnetic quantum critical point is predicted to yield significant information and should enable a better understanding of the problem in Ce-based noncentrosymmetric superconductors. From this standpoint, the noncentrosymmetric superconductor LaRhSi$_{3}$ is an ideal system for extensive investigations. We have therefore investigated LaRhSi$_{3}$ with the expectation that it will provide information to enrich our understanding of the relationship between the superconductivity and lack of symmetry in general. This will also provide comparative results for the superconducting state of CeRhSi$_{3}$.

The preliminary report based on resistivity measurements by Lejay et al. reveals an onset of superconductivity between 1.9 K and 2.7 K in LaRhSi$_{3}$  \cite{20}. This compound forms in the BaNiSn$_{3}$-type tetragonal structure (space group \textit{I4 mm}) in which Rh and Si atoms lack inversion symmetry along the c-axis. Recently de Haas-van Alphen (dHvA) studies have been carried out on single crystal LaRhSi$_{3}$, to investigate the Fermi surface properties  \cite{21,22} which, together with electronic structure calculations, predict the Fermi surface to consist of three asymmetry-split sheets.  dHvA results also show effective masses up to 1.6$m_{e}$ for different frequency branches, and spin-orbit coupling of the order of 10$^{2}$ K in LaRhSi$_{3}$. In this paper we present our results obtained from detailed investigations by neutron diffraction, magnetization, resistivity, specific heat and $\mu$SR measurements on LaRhSi$_{3}$ and characterize it as a moderate dirty-limit s-wave weakly coupled type-I superconductor with an isotropic superconducting gap and singlet pairing ground state.

\section*{Experimental}

The polycrystalline sample of LaRhSi$_{3}$ was prepared by the standard arc melting technique on a water cooled copper hearth under an inert argon atmosphere using the high purity elements (99.9\% and above) in stoichiometric ratio. To improve the homogeneity and reaction among the constituent elements, the sample was flipped several times during the melting process and subsequently annealed at 900 $^{o}$C for a week under a dynamic vacuum. The crystal structure was determined by Cu-K$_{\alpha}$ powder X-ray diffraction. The magnetic susceptibility was measured by a commercial SQUID Magnetometer (MPMS, Quantum-Design, San Diego) with an iQuantum $^{3}$He outfit (Quantum-Design, Japan). The specific heat was measured by the relaxation method in a PPMS (Quantum-Design, San Diego). The electrical resistivity was measured by the standard four probe ac technique using the PPMS. The $\mu$SR measurements were carried out using the MuSR spectrometer at the ISIS Facility at the Rutherford Appleton Laboratory, Didcot, U.K., both in longitudinal and transverse geometry. The powder sample was mounted on a silver holder (purity 4N) with GE-varnish to improve thermal equilibrium. The use of silver ensures a time independent background contribution to the $\mu$SR spectra as silver gives only a nonrelaxing muon signal. The stray fields at the sample position were cancelled to within 1 $\mu$T by using correction coils. The neutron diffraction experiment was performed on a powdered sample at 300K using the ROTAX diffractometer at the ISIS Facility.

\begin{figure}
\includegraphics[width=10.0cm, keepaspectratio]{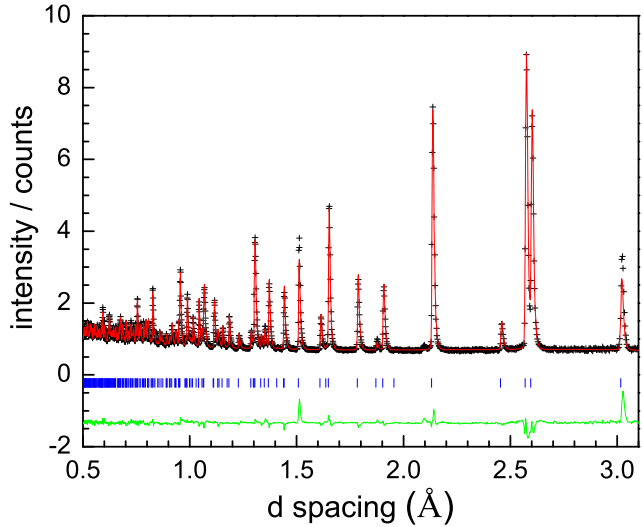}
\caption{\label{fig1} (colour online) Neutron diffraction pattern of LaRhSi$_{3}$ recorded at room temperature. The solid line through the experimental points is the Rietveld fit profile using the BaNiSn$_{3}$-type tetragonal structure (space group \textit{I4 mm}). The short vertical bars mark the theoretical Bragg diffraction positions. The lowermost curve represents the difference between the experimental and calculated results.}
\end{figure}

\section*{Results and Discussion}

The X-ray diffraction data collected from a powdered sample of LaRhSi$_{3}$ at room temperature were analysed by Rietveld refinement using Fullprof software. The crystal structure was confirmed to be BaNiSn$_{3}$-type tetragonal structure (space group \textit{I4 mm}) with lattice parameters $a$ = 4.2694(03) {\AA} and $c$ = 9.8357(10) {\AA}, in very good agreement with the literature value  \cite{20}. For the best fit using the least squares refinement method $\chi^{2}$ had the value of 1.24. No impurity phase was detected in powder X-ray diffraction data.  To characterise the whole bulk volume of the sample, we carried out a neutron diffraction study. Fig.1 shows our neutron diffraction pattern of LaRhSi$_{3}$ recorded at room temperature together with the structural Rietveld refinement profile using the GSAS software for the BaNiSn$_{3}$-type tetragonal structure (space group \textit{I4 mm}) model. The results obtained from a least squares refinement of neutron diffraction are listed in Table I. During the refinement the occupancy of all the elements was kept fixed as its variation was not improving the fit quality. The lattice parameters are in perfect agreement with those obtained from the powder X-ray diffraction, and the neutron results confirmed the single-phase nature of the bulk sample.

\begin{table}
\caption{\label{tab:table1} Crystallographic and refinement parameters of LaRhSi$_{3}$ determined from the full structural refinement of neutron diffraction data using the GSAS program.}
\begin{ruledtabular}
\begin{tabular}{llccc}

Structure & BaNiSn$_{3}$-type tetragonal\\

Space group & \textit{I4 mm} (No. 107)\\

f.u./unit cell &   2\\

Crystal parameters\\

  \hspace{1.0 cm}   $a$        &  4.2693(4) {\AA} \\

  \hspace{1.0 cm}    $c$       &   9.8292(9) {\AA} \\

  \hspace{1.0 cm}  $V_{cell}$  &   179.15(5) {\AA}$^{3}$ \\

  \hspace{1.0 cm}  $V_{mole}$  &   53.94 cm$^{3}$/mole \\

Refinement Quality & Parameters \\

  \hspace{1.0 cm}     $R_{p}$  & 3.03\% \\

  \hspace{1.0 cm}     $R_{wp}$ & 3.51\% \\

Atomic Coordinates \\

   \hspace{1.0 cm} Atom \hspace{0.33 cm} x  & y \hspace{1.0 cm} z \hspace{0.85 cm} Mult  & Occupancy  & $U_{iso}$ ({\AA}$^{2}$) \\

   \hspace{1.0 cm}    La  \hspace{0.77 cm} 0 & 0 \hspace{0.4 cm} 0.00265(11) \hspace{0.3 cm} 2 & 1.0 & 0.00680(34)\\

   \hspace{1.0 cm}    Rh  \hspace{0.73 cm} 0 & 0 \hspace{0.4 cm} 0.65771(13) \hspace{0.41 cm}2 & 1.0 & 0.00202(35)\\

   \hspace{1.0 cm}    Si1 \hspace{0.73 cm} 0 & 0 \hspace{0.4 cm} 0.41350(20) \hspace{0.3 cm} 2 & 1.0 & 0.00772(42) \\

   \hspace{1.0 cm}    Si2 \hspace{0.73 cm} 0 & 0.5 \hspace{0.15 cm} 0.26503(18)\hspace{0.42 cm} 4 & 1.0 & 0.01077(34) \\

\end{tabular}
\end{ruledtabular}
\end{table}

\begin{figure}
\includegraphics[width=10.0cm, keepaspectratio]{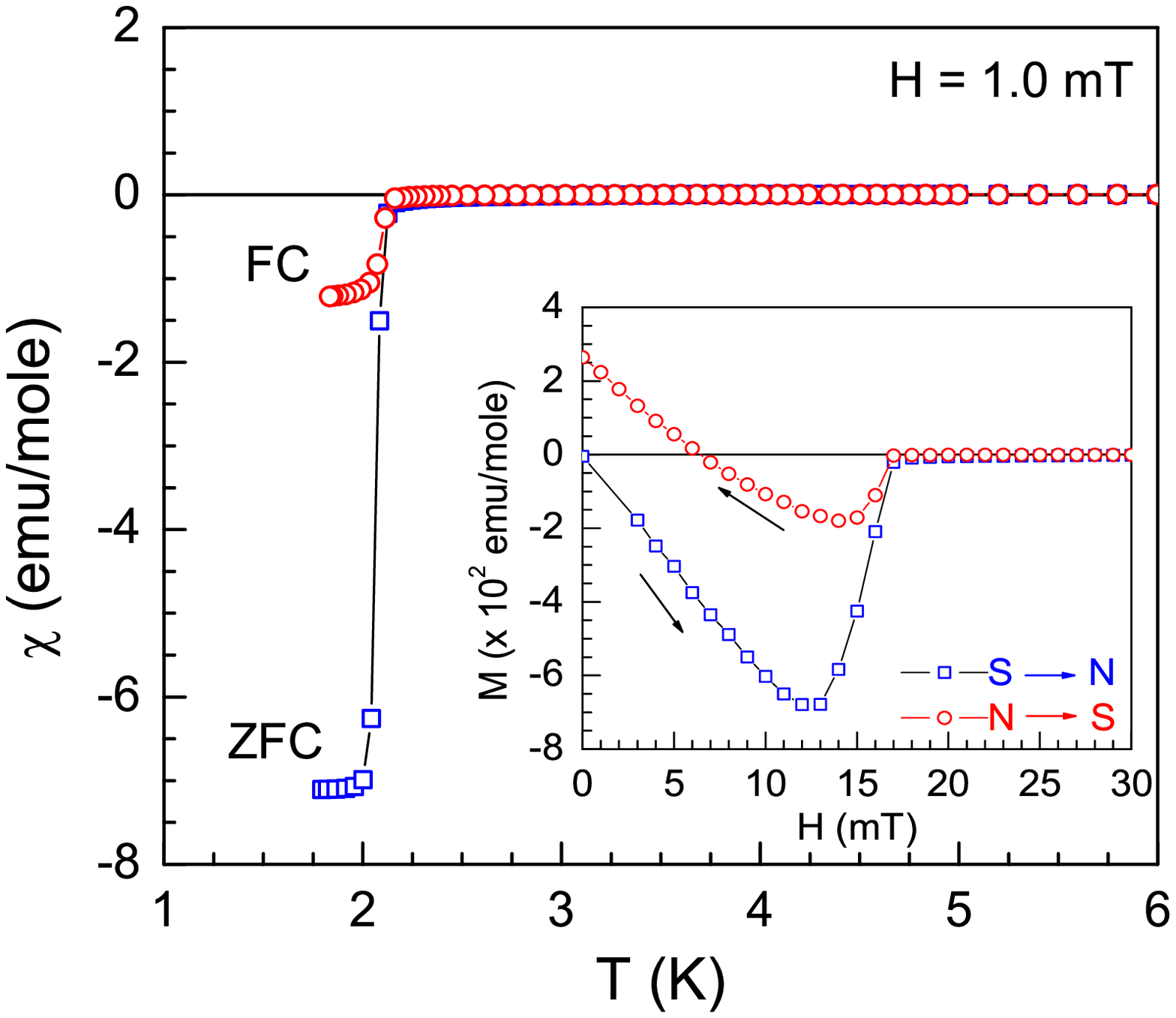}
\caption{\label{fig2} (colour online) Temperature dependence of low temperature zero field cooled (ZFC) and field cooled (FC) magnetic susceptibility $\chi(T)$ data of LaRhSi$_{3}$ measured at 1.0 mT. The inset shows the isothermal magnetization as a function of magnetic field measured at 0.5 K. Arrows indicate the directions for the magnetic field cycle between the normal (N) and superconducting (S) states.}
\end{figure}

Figure 2 shows the low temperature magnetic susceptibility $\chi(T)$ data measured at a field of 1.0 mT. Both the zero field cooled (ZFC) and field cooled (FC) $\chi(T)$ data exhibit a large Meissner signal below 2.2 K, demonstrating the onset of superconductivity ($T_{c}$ = 2.16 K). An estimate of the superconducting phase fraction using the zero field cooled magnetization data yields a Meissner volume fraction of $\sim$ 100($\pm$10) \%, indicating bulk superconductivity in this compound. The inset of Fig.2 shows the isothermal magnetization as a function of magnetic field measured at a constant temperature of 0.5 K. The hysteresis of the magnetization curve follows a near-typical type-I superconducting behaviour. The departure from the ideal step transition at critical field can be attributed to the geometrical shape effect of our sample (demagnetization factor). The temperature dependence of the thermodynamic critical field $H_{c}(T)$ determined from the low field magnetization measurements at different temperatures is shown in Fig.7, together with that determined from the specific heat data. $H_{c}(T)$ fits well to the relation $H_{c}(T) = H_{c0}[1 - (T/T_{c0})^{\alpha}]$ with the fitting parameters $H_{c0}$ = 18.1 $\pm$ 0.2 mT and $\alpha$ = 1.85 $\pm$ 0.06 using the value of $T_{c0}$ = 2.16 K. The value of $\alpha$ = 1.85 thus obtained is very close to the conventional value of $\alpha$ = 2. The $H_{c}$(T) data can also be fitted to the conventional relation with $\alpha$ = 2, i.e., $H_{c}(T) = H_{c0}^{*}[1 - (T/T_{c0})^{2}]$, the resulting parameter $H_{c0}^{*}$ being 17.6 $\pm$ 0.2 mT. However, the quality of fit is better with $\alpha$ = 1.85. Thus, from magnetization data, we estimate the thermodynamic critical field to be 18.1 mT subject to the correction due to the demagnetization factor.

\begin{figure}
\includegraphics[width=10.0cm, keepaspectratio]{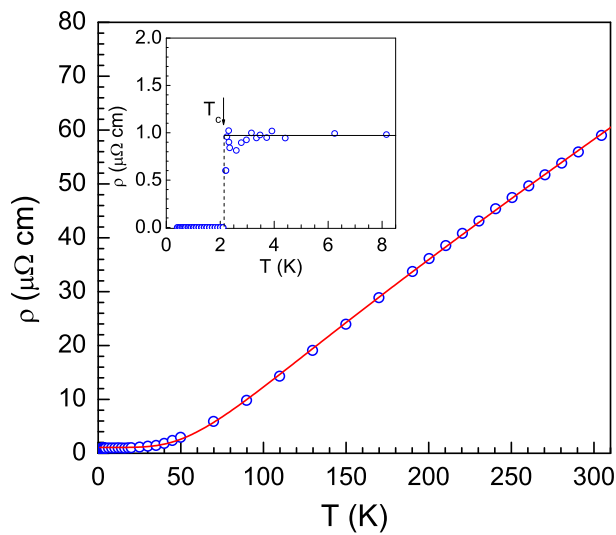}
\caption{\label{fig3} (colour online) Electrical resistivity of LaRhSi$_{3}$ as a function of temperature measured in zero magnetic field. The solid line represents our fit to the Bloch-Grüneisen model. The inset shows the expanded view of the low temperature data showing the superconducting transition. The lines are drawn as a guide to the eye.}
\end{figure}

Figure 3 shows the electrical resistivity data of LaRhSi$_{3}$ measured in zero field. While the high temperature resistivity exhibits metallic behaviour, at low temperature (despite the presence of noise) a sharp transition at 2.16 K (the transition mid-point, with an onset temperature 2.24 K) to a zero resistance state clearly indicates superconductivity in this compound. The normal state resistivity is well described by the Bloch-Gr\"{u}neisen model,

\begin{displaymath}
\rho(T) =\rho _0 + \frac {4B}{\theta_D} \left( \frac{T}{\theta _D} \right)^5 \int_0^{\theta _D} \frac{z^5 dz}{(e^z-1)(1-e^{-z})}
\end{displaymath}

\noindent where $\rho_0$ is the residual resistivity due to static defects in the crystal lattice and the spin-disorder resistivity due to the presence of disordered magnetic moments, and the second term represents the phonon assisted electron scattering ($\theta_{D}$ is the Debye temperature and $B$ is the electron-phonon coupling constant). A least-squares fitting of resistivity data above 2.5 K to this expression (solid line in Fig.3) gives $\rho_{0}$ = 1.08 $\mu \Omega$ cm, $B$ = 24.8 m$\Omega$ cm K and $\theta_{D}$ = 348 K. The low value of the residual resistivity $\rho_{0}$ of $\sim$ 1 $\mu \Omega$ cm just above the superconducting transition and a residual resistivity ratio of about 60 clearly reflect the good quality of our sample. This value of residual resistivity together with the electron carrier density can be used to estimate the mean free path, $ l = v_{F} \tau$, where the Fermi velocity $v_{F} = \hbar k_{F}/m^{*}$ and $\tau$ is the scattering time given by $\tau^{-1} = ne^{2}\rho_{0}/m^{*}$ for the Drude model. The effective mass $m^{*}$ as estimated from the relation for the electronic specific heat coefficient $\gamma = \pi^{2} n m^{*} k_{B}^{2}/ \hbar^{2}k_{F}^{2}$ turns out to be $m^{*}$ = 2.14 $m_{e}$ using $\gamma_{n}$ = 6 mJ/mole K$^{2}$ (as discussed in the following paragraphs) which is slightly larger than that observed in dHvA measurements ($m^{*} \sim$ 1.6 $m_{e}$) \cite{22}. Since the space group \textit{I4 mm} contains two formula units per unit cell, for our compound there are two La ions, each contributing three conduction electrons, and hence 6 conduction electrons per unit cell.  Therefore the electron density can be roughly estimated as $n = 6/V_{cell}$ = 3.349 $\times$ 10$^{28}$ m$^{-3}$. These values of $n$ and $m^{*}$ together with $\rho_{0}$ yield a Fermi velocity $v_{F}$ = 5.39 $\times$ 10$^{5}$ m/s and mean free path $l$ = 122 nm.

\begin{figure}
\includegraphics[width=10.0cm, keepaspectratio]{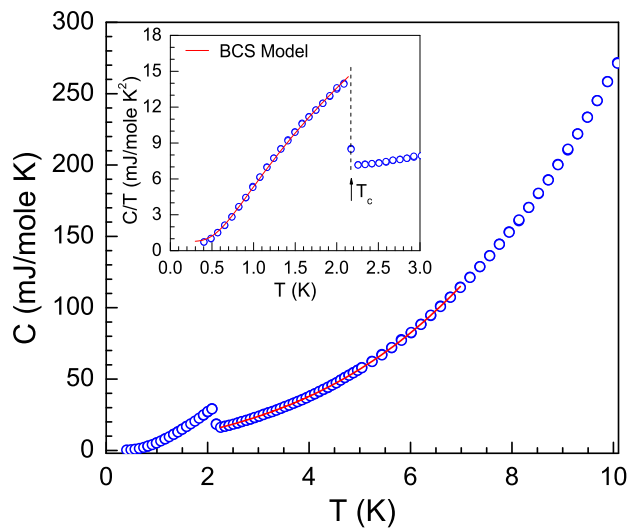}
\caption{\label{fig4} (colour online) Specific heat $C(T)$ data of LaRhSi$_{3}$ as a function of temperature measured in zero field. The solid line (above $T_{c}$) is a fit to $C = \gamma T + \beta T^{3}$. The inset shows the expanded view near the superconducting transition, plotted as $C/T$ \textit{vs.} $T$. The solid line in the inset represents the theoretical temperature-dependent spin-singlet fully gapped superconductor according to the weak coupling BCS model as tabulated by M\"{u}hlschlegel, Ref. 23.}
\end{figure}

Figure 4 shows the specific heat data of LaRhSi$_{3}$. A sharp transition in the specific heat confirms the intrinsic nature of superconductivity in this compound, in agreement with the magnetization measurements discussed above. We define the critical temperature as the approximate mid point of the transition, $T_{c}$ = 2.16 K. Above the transition temperature, i.e. in the normal state, the specific heat data is well represented by $C = \gamma T + \beta T^{3}$. A linear fit to $C/T$ \textit{vs.} $T^{2}$ plot in the temperature range 2.25 K to 7 K gives the Sommerfeld coefficient $\gamma_{n}$ $\approx$ 6.0 mJ/mole K$^{2}$ and $\beta$ $\approx$ 213.6 $\mu$J/mole K$^{4}$. From the value of $\beta$ we estimate the Debye temperature to be 357 K using the relation $\theta_{D} = (12 \pi^{4} N_{A} r k_{B} / 5 \beta)^{1/3}$, where $r$ is the number of atoms per formula unit, which is consistent with the $\theta_{D}$ value estimated from the resistivity data. Further, from the observed jump in the specific heat at $T_{c}$, $\Delta C_{el}$ = 16 mJ/mole K, the ratio $\Delta C_{el}/ \gamma_{n} T_{c}$ $\approx$ 1.25 which is comparable to 1.43, the BCS expected value in the weak coupling limit. The electronic specific heat coefficient in the superconducting state is estimated from the difference between the specific heats observed in the superconducting state in zero field and that under an applied magnetic field of 15.0 mT (a field of 15.0 mT suppresses the $T_{c}$ to below 0.45 K as will be discussed later), $\gamma_{s}$ $\approx$ 5.4 mJ/mole K$^{2}$ giving $\Delta C/ \gamma_{s} T_{c}$ $\approx$ 1.37 which is very close to the weak coupling BCS value of 1.43. Further, the values of $\gamma_{n}$ and $\gamma_{s}$ suggest a superconducting volume fraction of at least 90\% signifying the bulk nature of BCS superconductivity in LaRhSi$_{3}$. The BCS-type superconductivity in this compound also follows from the temperature dependence of the specific heat in the superconducting state. The experimentally observed data in the superconducting state could be reasonably reproduced by the generalised weak-coupling BCS dataset of M\"{u}hlschlegel  \cite{23} (solid line in the inset of Fig. 4) suggesting a fully gapped spin-singlet BCS superconductivity in this compound. To achieve a better agreement between the theoretical M\"{u}hlschlegel dataset and experimentally observed data we have adjusted the values of $\gamma$ and $T_{c}$. The solid line in the inset of Fig.4 corresponds to $\gamma^{*}$ = 5.28 mJ/mole K$^{2}$ (which is close to our $\gamma_{s}$ value for the superconducting state) and $T_{c}^{*}$= 2.14 K, which gives us a thermodynamic mean value of the critical temperature. Even though the specific heat data are well interpreted with a full BCS gap by adjusting $\gamma$ and $T_{c}$, to estimate the superconducting energy gap more precisely we analyze the electronic part of the specific heat in the superconducting state (plotted in Fig.5) which is obtained from the difference between the specific heat data measured in zero field and that measured in 15.0  mT, i.e. $C_{el}(T) = \Delta C(T) = C(T)_{0} - C(T)_{15}$. As expected for the BCS ground state, the electronic part of the specific heat $C_{el}$ below $T_{c}$ exhibits an exponential temperature dependence, confirming the s-wave pairing. The solid line in Fig.5 represents the fit to $C_{el}(T) \sim T^{1/2} exp(-\Delta_{0}/k_{B}T)$, with an energy gap of $\Delta_{0}$ = 3.50 $\pm$ 0.06 K ($\sim$ 0.3 meV). This gives $2 \Delta_{0} /k_{B} T_{c}$ = 3.24 which is in reasonable agreement with the weak-coupling BCS expected value of 3.52. An estimate of the superconducting gap $\Delta_{0}$ from the relation $\mu_{0} H_{c0}^{2} = (3\gamma /2 \pi^{2}k_{B}^{2})\Delta_{0}^{2}$ gives $\Delta_{0}$ = 3.73 K (using $H_{c0}$ = 17.2 mT) which is equivalent to $2\Delta_{0}/k_{B}T_{c}$ = 3.45, in better agreement with the BCS value. We thus see that the specific heat data provide compelling evidence for an s-wave isotropic BCS superconducting gap in the electronic density of states right at the Fermi energy level.

\begin{figure}
\includegraphics[width=10.0cm, keepaspectratio]{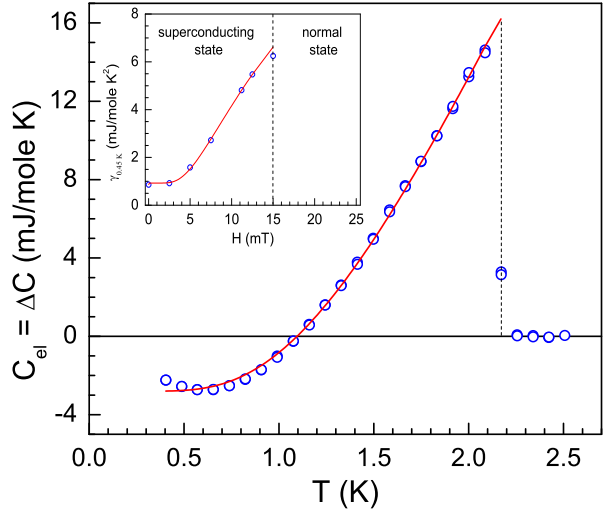}
\caption{\label{fig5} (colour online) Temperature dependence of the electronic part of the specific heat, C$_{el}$(T) of LaRhSi$_{3}$, the solid line is the fit assuming an isotropic s-wave BCS superconducting gap, $C_{el}(T) \sim T^{1/2} exp(-\Delta_{0}/k_{B}T)$. The inset shows the field dependence of $\gamma$, where the solid line represents an exponential evolution of $\gamma$ under the application of magnetic field.}
\end{figure}

We also measured the specific heat of LaRhSi$_{3}$ under the application of selected magnetic fields of 2.5, 3.5, 5.0, 7.5, 11.2, 12.5 and 15.0 mT (Fig.6) to see the effect of a magnetic field on the transition temperature and obtain information on the temperature dependence of the critical field. As seen from Fig.6 the superconducting transition temperature, $T_{c}$, decreases rapidly with the application of field, e.g. at a field of 5.0 mT, $T_{c}$ is reduced to 1.62 K from its value of 2.16 K at zero field, and superconductivity is suppressed to below 0.45 K at a field of 15.0 mT. To see the evolution of $\gamma$ with the magnetic field we plot $\gamma(H) = C(T)_{H}/T$ at 0.45 K as a function of magnetic field (see inset of Fig. 5).  The experimentally observed data exhibit an exponential field dependence, $\gamma(H) \sim exp(-H^{*}/H)$ with $H^{*}$ $\approx$ 17 mT which is similar in magnitude to the thermodynamic critical field $H_{c0}$. This clearly suggests that $\gamma$, and hence the nonsuperconducting density of states, evolves exponentially with magnetic field. For an isotropic gapped superconductor one would expect a linear field dependence of $\gamma$(H). We suspect that the superconducting gap which is isotropic in zero field becomes anisotropic with the field, and the anisotropy gets stronger with increasing field, which would then imply that the mechanism for superconductivity in LaRhSi$_{3}$ may be different from the conventional BCS picture. Another interesting feature observed in the specific heat data under the application of magnetic field is that the jump in specific heat at the transition is larger for 2.5 mT than that for zero magnetic field (Fig.6), the characteristic of a first-order transition. That the application of magnetic field drives the superconducting transition from second-order in zero magnetic field to a first-order transition in nonzero magnetic fields strongly suggests a type-I superconductivity in this compound.

\begin{figure}
\includegraphics[width=10.0cm, keepaspectratio]{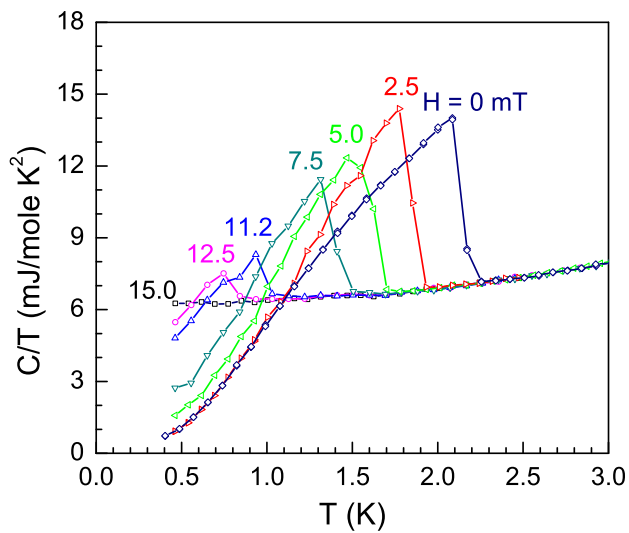}
\caption{\label{fig6} (colour online) Temperature dependent specific heat $C(T)_{H}$ data below 3 K measured under the application of different fields ranging from 0 to 15.0 mT, plotted as $C/T$ \textit{vs.} $T$.}
\end{figure}

\begin{figure}
\includegraphics[width=10.0cm, keepaspectratio]{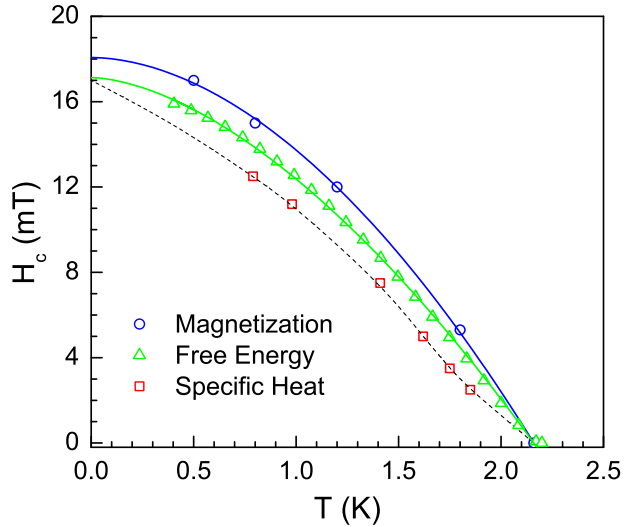}
\caption{\label{fig7} (colour online) Temperature dependence of thermodynamic critical field $H_{c}(T)$ for LaRhSi$_{3}$ determined from the low field magnetization measurements at different temperatures and specific heat measurement under the application of magnetic field as well as that calculated from free energy considerations from zero field specific heat data. The solid lines represent the parabolic fit to $H_{c}(T) = H_{c0}[1 - (T/T_{c0})^{\alpha}]$ as discussed in the text. The dashed line is a guide to the eye (following a polynomial behaviour for $T < 0.8 T_{c}$)}
\end{figure}

In Fig.7 we have plotted the magnetic field \textit{vs}. temperature, $H-T$ phase diagram determined from the field dependence of the superconducting transition temperature, $T_{c}$, as obtained from the specific heat measurements in an applied magnetic field. The apparent upward curvature near $T_{c0}$ in $H(T)$ for the case of the specific heat measurement under magnetic field in Fig.7 can be attributed to the effect of the demagnetizing field. The magnitude of the latter effect is clearly visible on the isothermal magnetization $M(H)$ displayed in the inset of Fig.2 and corresponds approximately to the spread of critical fields from the three data sets displayed in Fig.7. We estimate the thermodynamic critical field $H_{c0}$ of LaRhSi$_{3}$ using the zero field specific heat data by integrating the entropy difference between the superconducting and normal states, i.e., by $\Delta F(T) = F_{n}(T)-F_{s}(T) = H_{c}^{2}(T)/8 \pi  = \int_{T_{c}}^{T}\int_{T_{c}}^{T'} \frac{C_n-C_s} {T''} dT'' dT' $,  where $F_{n}$ and $F_{s}$ are the free energies per unit volume in the normal and superconducting states respectively. $H_{c}(T)$ obtained from free energy considerations is also shown in Fig.7. The critical field $H_{c0}$ is obtained from a fit to the conventional relation $H_{c}(T) = H_{c0}[1 - (T/T_{c0})^{\alpha}]$ below 2.1 K only to avoid the error due to the curvature near $T_{c0}$. The best fit gives $H_{c0}$ = 17.1 $\pm$ 0.1 mT and $\alpha$ = 1.66 $\pm$ 0.02. The value of $\alpha$ is slightly reduced compared to what we have deduced from the magnetization ($\alpha$ = 1.85). Thus from free energy calculations we obtain the thermodynamic critical field $H_{c0}$ = 17.1 mT.

An estimate of the upper critical field following the WHH approach  \cite{24}] for a conventional type-II superconductor, which predicts $H_{c2} \approx 0.69(dH_{c2}(T)/dT)T_{c0}$, yields $H_{c2} \approx$ 17 mT using the slope of $H_{c}(T)$ in the temperature range $0.6 T_{c} <  T <  0.8 T_{c}$, $dH_{c}(T)/dT$ = 11.4 mT/K and $T_{c0}$ = 2.16 K. This value of the upper critical field is very close to the thermodynamic critical field estimated from free energy considerations, implying a type-I behaviour in LaRhSi$_{3}$. The WHH model, which estimates the critical field in terms of orbital pair breaking, accounts for  both spin-orbit scattering and Pauli spin paramagnetism (or Maki parameter). The Pauli paramagnetic limiting field corresponds to the field at which $F_{n}(H)$ equals the condensation energy of the superconducting state, and, for the weak coupling case, the Pauli-Clogston limiting field is given by $H_{P} = 1.86~T_{c}$  \cite{25,26}. The Pauli-limiting field of 4.02 T for LaRhSi$_{3}$ is very high compared to the estimated field of 17.1 mT, suggesting the absence of a Pauli limiting field in $H_{c}$ of this compound. The value of the Maki parameter $\alpha$, which provides information about the relative strength of orbital and spin pair-breaking, can be estimated from the Sommerfeld coefficient $\gamma_{n}$ and residual resistivity $\rho_{0}$, $\alpha = (3e^{2}\hbar \gamma \rho_{0})/(2m\pi^{2} k_{B}^{2})$  \cite{24,27}, giving $\alpha$ = 0.003. Alternatively, using the slope of the $H_{c}(T)$ curve, $\alpha = 5.2758 \times 10^{-5} (\frac {dH_{c}(T)}{dT})|_{T=T_c}$ = 0.006, which is the same order of magnitude as the above estimated value. The value of $\alpha$ obtained for LaRhSi$_{3}$ is clearly very low, suggesting that the critical field is essentially determined by the orbital pair breaking.

The electron-phonon coupling $\lambda_{e-ph }$, which determines the attractive part of the Cooper pair bonding, was estimated using the value of $\theta_{D}$ and T$_{c}$ following McMillan's theory  \cite{28},

\begin{displaymath}
 \lambda_{e-ph}= \frac {1.04+\mu^{*} ln(\theta_D/1.45T_c)} {(1-0.62\mu^{*})ln(\theta_D/1.45T_c) - 1.04}
\end{displaymath}

\noindent where $\mu^{*}$ represents the repulsive screened Coulomb part, which is usually taken between 0.1 and 0.15. Setting $\mu^{*}$ = 0.13, $\lambda_{e-ph}$ for our compound comes out to be $\approx$ 0.5 which implies LaRhSi$_{3}$ is a weak coupling superconductor.

The coherence length in the clean limit is obtained by the BCS relation $\xi_{0} = 0.18\hbar v_{F} / k_{B} T_{c}$, which gives $\xi_{0}$ = 343 nm. Alternatively, the coherence length for $T \rightarrow 0$ can be estimated by using the relation $\xi_{0} = 7.95 \times 10^{-17} [n^{2/3 }(S/S_{F})](\gamma T_{c})^{ -1}$ cm, where $n$ is the conduction electron density in units of cm$^{-3}$, $\gamma$ is expressed in erg/cm$^{3}$ K$^{2}$ and $S/S_{F}$ is the ratio of the Fermi surface area ($S$) of the superconducting electron density, to  the Fermi surface ($S_{F}$) of the free electron gas density $n$  \cite{29}. Assuming a simple model of a spherical Fermi surface ($S/S_{F}$ = 1) we obtain $\xi_{0}^{*}$ = 344 nm, similar to the above value. Within this approach we can also estimate the mean free path from the relation $l_{tr} = 1.27 \times 10^{4} [\rho_{0} n^{2/3 }(S/S_F)]^{-1}$, $\rho_{0}$ being in $\Omega$~cm, and for $S/S_{F}$ = 1 we obtain $l_{tr}$ = 122 nm which is precisely the same as obtained above within the Drude model. It is clearly inferred that the mean free path is considerably smaller than the BCS coherence length ($l/\xi_{0}$ $\approx$ 0.36) which in turn suggests that LaRhSi$_{3}$ can be classified as a moderately dirty-limit superconductor. The estimated value of Gorkov's impurity parameter, $\alpha_G$ = 2.5, further supports this classification.

The London penetration depth estimated from $\lambda_{L}^{2} = m^{*}c^{2}/4 \pi n e^{2}$ comes out to be 43 nm, which is in good agreement with the alternative estimate from $\lambda_{L} = 1.33 \times 10^{8} \gamma^{1/2} [n^{2/3 }(S/S_F)]^{-1}$ giving $\lambda_{L}^{*}$ = 44 nm for $S/S_{F}$ = 1. The ratio $\lambda_{L}/\xi_{0}$ = 0.12 $< 1/\surd2$, clearly classifying LaRhSi$_{3}$ as a type-I superconductor. Further, using the relation for the Ginzburg-Landau parameter $\kappa = 7.49 \times 10^{3} \gamma^{1/2} \rho_{0}$ for a dirty-limit superconductor, we get $\kappa$ = 0.25, consistent with the type-I superconductivity in LaRhSi$_{3}$. In the dirty limit, the Ginzburg-Landau coherence length can be obtained from the relation $\xi_{GL} = 8.57\times 10^{-7}(\gamma \rho_{0} T_{c})^{-1/2}$. This gives $\xi_{GL}$ = 175 nm, which in turn from the definition $\kappa = \lambda_{GL}/ \xi_{GL}$, gives a Ginzburg-Landau penetration depth $\lambda_{GL}$ = 44 nm.

The enhanced density of states is found from the relation $N^{*}(E_{F}) = 0.2121 \gamma/N$, where $N$ is the number of atoms per formula unit and $\gamma$ is expressed in mJ/mole K$^{2}$, which gives $N^{*}(E_{F})$ = 0.25 states/[eV atom spin-direction]. The bare density of states, given by $N(E_{F}) =N^{*}(E_{F})/(1+\lambda_{e-ph})$, is 0.17 states/[eV atom spin-direction]. The measured and derived superconducting parameters of LaRhSi$_{3}$ are listed in Table II. In deriving the various superconducting parameters we have assumed a spherical Fermi surface. To verify the self consistency of our assumption, we evaluate the electronic coefficient of specific heat, $\gamma$, from the thermodynamic critical field, using the relation $\gamma = 2.12 \mu_{0}H_{c0}^{2}/T_{c0}^{2}$. Taking $H_{c0}$ = 17.2 mT as obtained from the $\mu$SR measurements (discussed in the following paragraphs), we obtain $\gamma_{es}$ = 5.76 mJ/mole K$^{2}$ which is very close to the experimentally observed value of $\gamma_{ob}$ = 6.04 mJ/mole K$^{2}$. This agreement between the electronic specific heat coefficient derived from the thermodynamic critical field and that observed experimentally validates our assumption of a spherical Fermi surface. Therefore we can safely say that the error introduced on account of the shape of Fermi surface in deriving the parameters listed in Table II must be small, and does not affect our conclusions of the essential physics deduced from our data. However to obtain the precise values of the derived parameters, one would need to have a better estimate of the electron density n, such as by Hall-effect measurements. Given that the band structure calculations for LaRhSi$_{3}$ clearly reveal a band splitting due to the noncentrosymmetric structure and spin-orbit coupling [22], one would expect that the superconducting properties will be dictated by antisymmetric spin-orbit coupling (ASOC). However, it seems that it is not strong enough to demonstrate its effect on the superconducting properties of LaRhSi$_{3}$, as is the case with CeRhSi$_{3}$, CeIrSi$_{3}$ and CePt$_{3}$Si. The reinforcement of ASOC with magnetic field might be responsible for the exponential evolution of $\gamma$ with magnetic field causing a field dependent anisotropic order parameter in the superconducting state.

\begin{table}
\caption{\label{tab:table2} Measured and derived superconducting parameters of the noncentrosymmetric superconductor LaRhSi$_{3}$.}
\begin{ruledtabular}
\begin{tabular}{lll}

$T_{c}$ (K) & 2.16 $\pm$ 0.08 \\

$H_{c}$ (mT) & 18.1 $\pm$ 0.2 --- magnetization \\

          {} & 17.1 $\pm$ 0.1 --- specific heat \\

          {} & 17.2 $\pm$ 0.1 --- $\mu$SR \\

$\gamma_{n}$ (mJ/mole K$^{2}$) & 6.04 $\pm$ 0.01 \\

$\beta$ ($\mu$J/mole K$^{4}$) &  213.63 $\pm$ 0.04 \\

$\theta_{D}$ (K) & 357 \\

$\gamma_{s}$ (mJ/mole K$^{2}$) & 5.4 \\

$\Delta C_{el} / \gamma_{n} T_{c}$ & 1.25 \\

$\Delta C_{el} / \gamma_{s} T_{c}$ & 1.37 \\

$2\Delta_{0} / k_{B}T_{c}$ & 3.24 \\

$m^{*}$ & 2.14 m$_{e}$ \\

$k_{F}$ (nm$^{-1}$) & 9.97 \\

$v_{F}$ (m/s) & 5.39 $\times$ 10$^{5}$ \\

$E_{F}$ (eV) & 3.54 \\

$\lambda_{e-ph}$ & 0.499 \\

$\xi_{0}$ (nm) & 343 \\

$l$ (nm)& 122 \\

$\lambda_{L}$ (nm)& 43 \\

$\xi_{GL}$(0) (nm) & 175 \\

$\lambda_{GL}$(0) (nm) & 44 \\

$\kappa$ & 0.25 \\

$N^{*}(E_{F})$ & 0.25 states/eV atom spin \\

$N(E_{F})$ & 0.17 states/eV atom spin \\

\end{tabular}
\end{ruledtabular}
\end{table}

\begin{figure}
\includegraphics[width=10.0cm, keepaspectratio]{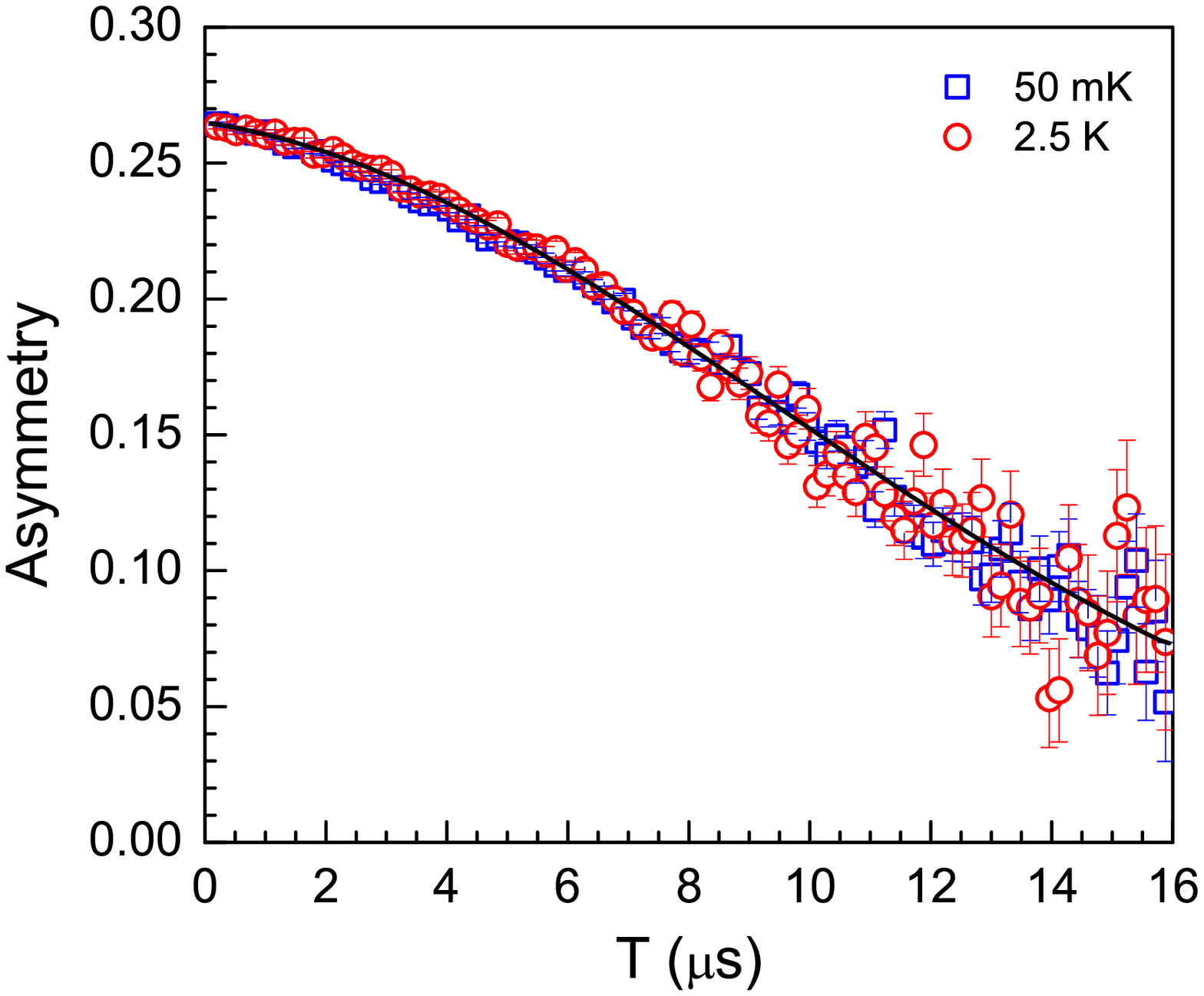}
\caption{\label{fig8} (colour online) Zero field $\mu$SR spectra measured in longitudinal geometry below (50 mK, squares) and above (2.5 K, circles) the superconducting transition temperature. The solid line is the fit to the Gaussian Kubo-Toyabe function as described in the text.}
\end{figure}

In order to further characterize the nature of the superconducting ground state of LaRhSi$_{3}$ we have used both muon spin relaxation and rotation measurements. Muon spin relaxation measurements were carried out in zero-field (longitudinal geometry) to investigate whether time reversal symmetry is broken as has been seen in the noncentrosymmetric superconductor LaNiC$_{2}$  \cite{10} as well as in a transverse field to characterise the superconducting ground state by estimating characteristic parameters. Our ZF $\mu$SR results above and below $T_{c}$ do not reveal any noticeable change in the relaxation rate (see Fig.8), which indicates the absence (within the sensitivity of $\mu$SR) of a spontaneous internal field at the muon site when entering the superconducting state. This confirms the preservation of time reversal symmetry when entering the superconducting state of LaRhSi$_{3}$.

The time evolution of muon polarization in zero field is best described by the Gaussian Kubo-Toyabe function,

\begin{equation}
 G_z(t)=A_0 \left(\frac{1}{3}+\frac{2}{3}(1-\sigma^2 t^2 )exp \left(-\frac{\sigma ^2 t^2}{2}\right)\right)exp(-\lambda t)+A_{bck}
\end{equation}

\noindent where $\sigma / \gamma_{\mu}$ is the local field distribution width, $\gamma_{\mu}$ = 13.553 MHz/T being the muon gyromagnetic ratio, and $\lambda$ is the electronic relaxation rate, A$_{0}$ is the initial symmetry and A$_{bck}$ is the background. The best fit was obtained for $\sigma$ = 0.067(3) s$^{-1}$ representing the random local field from nuclear moments, and a relaxation rate due to the electronic moments $\lambda$ = 0.013(4) s$^{-1}$.

The transverse field muon spin rotation data were collected after cooling the sample in an applied field from the normal state into the superconducting state. In Fig.9 we show the $\mu$SR spectra for applied magnetic fields of 5.0 and 15.0 mT both below ($T$ = 0.2 K) and above ($T$ = 2.5 K) the transition temperature. It is to be noted that the spectra in each of the detectors were decomposed into real and imaginary components: here we show only the real components. The spectra in 5.0 mT above $T_{c}$ reveals full initial asymmetry, while below $T_{c}$ there is a considerable reduction in the initial asymmetry. Further the spectra in 5.0 mT can be described using a single Gaussian oscillatory component, which gives a very similar frequency above and below $T_{c}$. The loss of initial asymmetry as observed in our $\mu$SR spectra of LaRhSi$_{3}$ can be compared with that of LaNiSn, which also exhibits type-I superconductivity  \cite{30}. In LaNiSn there is also a considerable reduction in the initial asymmetry at lower applied fields and then the asymmetry recovers in higher applied field, which is very similar to what we have seen in LaRhSi$_{3}$. However, if we use a different grouping method [($F-\alpha B$)/($F+ \alpha B)$, where F and B are the forward and backward detectors and $\alpha$ is a calibration constant] to analyze the 5.0 mT spectra at 0.2 K, we observe an offset in asymmetry instead (see inset of Fig.9(b)).

\begin{figure}
\includegraphics[width=10.0cm, keepaspectratio]{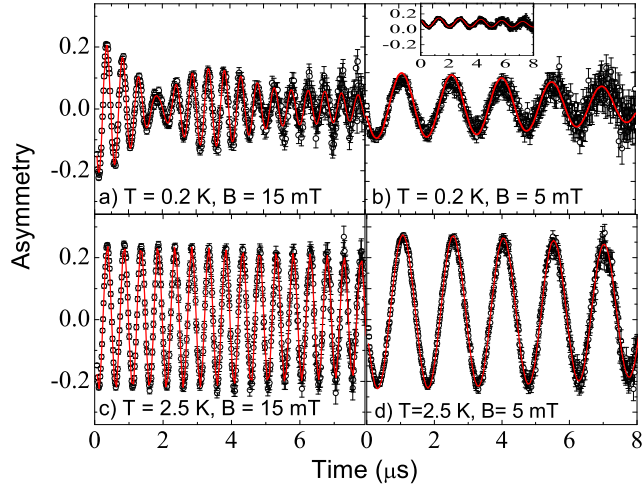}
\caption{\label{fig9} (colour online) The transverse field $\mu$SR spin precession signals recorded in transverse applied magnetic fields at (a) 0.2 K and 15 mT (intermediate state) and (b) 0.2 K and 5 mT (Meissner state)  (c) 2.5 K and 15 mT (normal state), and (d) 2.5 K and 5 mT (normal state). The solid lines are the fit to two oscillatory damped Gaussian functions for 15 mT (only one oscillatory function for 5.0 mT) as described in the text. The inset in (b) shows the 0.2 K and 5 mT spectra analysed by the different method discussed in the text.}
\end{figure}

\begin{figure}
\includegraphics[width=10.0cm, keepaspectratio]{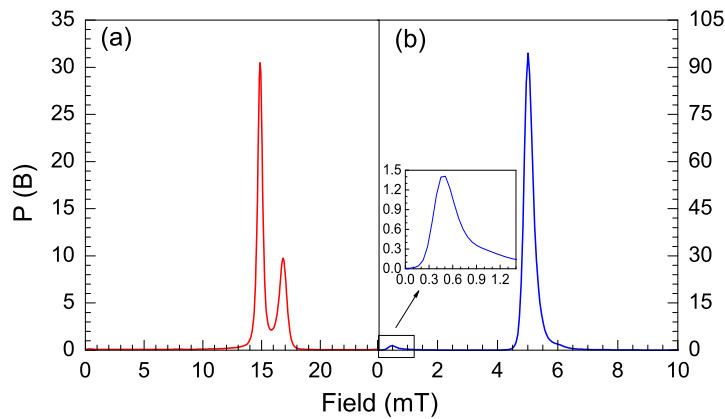}
\caption{\label{fig10} (colour online) The maximum entropy spectra for (a) 15.0 mT at 0.2 K, and (b) 5.0 mT at 0.2 K. Inset in (b) shows an expanded view to show the increase in P(B) near B = 0.}
\end{figure}

On the other hand the spectra in a 15.0 mT field clearly reveal the presence of two oscillatory terms. The spectra are best described by two oscillatory functions each damped with a Gaussian, i.e

\begin{equation}
 Gz(t)=\sum_{i=1}^{2} A_i cos(\omega_i t + \varphi) exp \left(-\frac{\sigma_i^2 t^2}{2}\right)
\end{equation}

\noindent where $A_{i}$ is the partial asymmetry ($A_{1}+A_{2} = A$), $\sigma_{i}$ is the relaxation rate, and $\omega_{i} = \gamma_{\mu}H_{i}$ is the central frequency for the respective components, $\gamma_{\mu}$ being the gyromagnetic ratio. Solid lines in the spectra show the best fit with this model, the fit parameters (for spectra at 0.2 K) are $A_{1}$ = 0.115, $\sigma_{1}$ = 0.02 $\mu$s$^{-1}$ and $\omega_{1}$ = 2.02 MHz for component 1 and $A_{2}$ = 0.096, $\sigma_{2}$ = 0.08 $\mu$s$^{-1}$ and $\omega_{2}$ = 2.28 MHz for component 2. From these parameters, we obtain the value of the internal magnetic field and weight fraction, which are 15.0 mT and 54.5 \% for the slow component and 17.2 mT and 45.5 \% for the fast component. The former value of the field is the same as the applied field (from the Silver holder), while the latter value can be taken as an estimate of the critical field coming from the intermediate state of the superconducting fraction of the sample for type-I behaviour.

It is worth to mention here that normally one would expect the $\mu$SR spectra to show the Kubo-Toyabe behaviour associated with nuclear fields, however, the data reduction used which rotates the spectra, effectively removes this contribution. We have also analyzed the $\mu$SR data of 5 mT at 0.2 K with a different method as mentioned above and the resultant spectra are shown in the inset of Fig.9(b). With this method an offset is observed in $\mu$SR asymmetry for T = 0.2 K, B = 5 mT (Meissner state). The spectra in the inset of Fig.9(b) were fitted using the sum of equations (1) and (2), but with only one component in equation (2). The decay is very weak because of the small nuclear moments. The quality of the fit can be seen from the figure. The maximum entropy spectra for the 5 mT data at 0.2 K is shown in Fig.10. As is expected for a sample in Meissner state, we observe an increase in P(B) near B = 0. However, we do not see increase in P(B) near B = 0 from Meissner volume in the intermediate state (for T = 0.2 K, B = 15 mT) of sample. We suspect that it is due to the effect of demagnetizing field, which is significant for a polycrystalline sample.

In Fig.10 we have also plotted the maximum entropy spectra for 15.0 mT at 0.2 K. Two sharp peaks in the maximum entropy spectra clearly demonstrate that the two oscillatory components of our model are at significantly different frequencies, one at $\sim$ 15.0 mT and the other at a somewhat higher value of $\sim$ 17.2 mT. For a type-II superconductor the establishment of a flux line lattice shows that the average field shifts to a lower frequency  \cite{31}, whereas for a type-I superconductor while entering the intermediate state (the regions of the sample are partially in normal and partially in superconducting states) the regions which are normal have an internal field which is equivalent to critical field. Thus $\mu$SR data give us a thermodynamic critical field $H_{c0}$ = 17.2 $\pm$ 0.1 mT in good agreement with the magnetization and specific heat data.

\section*{Conclusion}

We have examined the physical properties of the noncentrosymmetric superconductor LaRhSi$_{3}$ by detailed magnetization, specific heat and electrical resistivity measurements and found a sharp superconducting transition at $T_{c}$ = 2.16 K. While the zero field specific heat data provide evidence of bulk BCS superconductivity in a weak coupling regime, the specific heat data measured under the applied magnetic field strongly reflect a type-I superconductivity in this compound, as is also revealed by the field dependence of the magnetization. Superconducting parameters estimated within the framework of BCS theory from the electronic specific heat coefficient and residual resistivity not only provide conclusive evidence of type-I superconductivity, but also specify that LaRhSi$_{3}$ is a moderately dirty-limit superconductor. The microscopic study of superconductivity in LaRhSi$_{3}$ using $\mu$SR confirms conventional s-wave singlet pairing and a type-I superconductivity with a thermodynamic critical field of 17.2 mT. An antisymmetric spin-orbit coupling that is not sizable enough to dictate the superconducting properties in zero field, becomes reinforced in magnetic field leading to an exponential evolution of $\gamma$ with magnetic field and a field dependent anisotropic order parameter in the superconducting state is speculated.  Further investigations, preferably on single crystals of LaRhSi$_{3}$, would be highly desirable to understand better the microscopic details of the superconductivity in this compound.

\section*{Acknowledgements}

We would like to thank Dr W. Kockelmann for his assistance in collecting the neutron diffraction data on ROTAX at the ISIS Facility and Dr F. Pratt for interesting discussion. Authors VKA, ADH and DTA would like to acknowledge financial assistance from CMPC-STFC grant number CMPC-09108. AMS acknowledges financial assistance from the SA-NRF (grant 2072956).

\end{document}